\documentclass[twocolumn,secnumarabic,amssymb,nobibnotes,aps,prd,showpacs]{{revtex4-1}}
\usepackage{amsmath}
\usepackage{amssymb}
\usepackage[dvipdfm]{epsfig}
\usepackage{graphicx}

\begin{document}
\title {
Hierarchy of Frustrations as Supplementary Indices in Complex System Dynamics, Applied to the U.S. Intermarket}%
\author{Krzysztof Sokalski}%
\email[e-mail: ]{sokalski\_krzysztof@o2.pl}
\affiliation{Institute of Computer Science, 
Cz\c{e}stochowa University of Technology,
Al. Armii Krajowej 17, 42-200 Cz\c{e}stochowa, Poland}
\date{}%

\begin{abstract}
 A definition of frustration is expressed by transitivity of binary entanglement relation in considered complex system. Extending this definition into n-ary relation a hierarchy of  frustrations' notions is derived. As a complex system the U.S. Intermarket is chosen where the correlation coefficient of Intermarket indices' sectors play the role of entanglement's measure. In each hierarchy level
the frustration and the transitivity are interpreted as values of an order's measure for corresponding subsystem. The derived theory is applied to 1983-2012 data of the U.S. Intermarket.  
\end{abstract}
\maketitle

\section*{Introduction}
Frustration represents  situation where several optimization conditions compete with each other so that a system can not satisfy them simultaneously. Frustrated systems are characterized by the presence of metastable states among which the system "hesitates" to choose. These metastable states often change their order of stability as a function of external parameters to exhibit phase transitions from one state to another \citep{bib:KAWAMURA}. In the 21st century, the research of frustrations has received a revived interest and  new areas. 
Frustrations appear in different systems and different scales. In some systems such as modern magnetic materials and superconductors the frustrations are sources of their expected properties \citep{bib:Cond.Matt} The frustrations are observed also in nature phenomena on the level of molecular scales \citep{bib:bryng}. The most recent discovers of frustrations have been done in Markets. In these systems the frustrations  play crucial role in creation of realistic market's models \citep{bib:ahlg}. In this paper we introduce our own interpretation of the frustration in a plaquette consisting of complex system's components. Definition of frustration is expressed by transitivity of the correlation relation in considered system. Extending this definition into transitivity of the $m-$ary relation \citep{bib:pickett},\citep{bib:usan},\citep{bib:cristea} we create a hierarchy of frustrations describing the whole complex system consisting of the $m$ components and its subsystems consisting of $m-1,m-2,\dots 3,2$ components, respectively.  Notion of the transitivity and frustration are opposite magnitudes of an attribute characterizing $n$ bodies correlation in considered subsystems of the hierarchy, where $n=m,m-1,m-2,\dots,3,2$. In order to perform investigations of this hierarchy with respect to the transitivity and frustration we need an appropriate complex system and empiric data for its members.  The paper is organized in the following way. Section \ref{data} approaches both system and data. In Section \ref{defin} using the notion of the transitivity we define frustration on different levels of hierarchy using notion of the transitivity. So far the transitive and frustrated subsystems are labelled by the two numbers +1  and -1, respectively. In order to make them more realistic we introduce in Section \ref{measure} extended measures of transitivity varying in the continuous domain $[ -1, +1]$. Section \ref{ord} presents interpretation of transitivity as ordering relation. finally, in Section \ref{anal} using results of Sections \ref{defin}  and \ref{measure} we analyse the 1983-2012 data of the U.S. Intermarket with respect to transitivity and frustration.
\section{1983-2012 Empiric Data of the Considered U.S. Intermarket}\label{data}
An appropriate complex system should satisfy the following conditions:\\
1) The system and its subsystems produce data according to probability-based regime,\\
2) Produced data should be homogenous, reliable and complete,\\
3) The data produced by different subsystems should be time synchronous.\\
We would also like to have data which are easy to get and cheap. Following Murphy \citep{bib:murp} we complete his U. S. Intermarket with Gold. In this way we choose system which satisfies our requirements.  All considerations being done here base on the data supplied by \citep{bib:NICK}. 
On basis of the 1987 Crash's data of the U. S.  markets, Murphy
 derived the concept of the Intermarket Technical Analysis
involving four sectors.
Before the Murphy invention many people applied very simple technical analysis, such
as program trading and portfolio insurance. Such simple analysis could not predict the 
forthcoming stock collapse.
The events of 1987 provide a textbook example of how the
intermarket scenario works and make a compelling argument as to why stock market
participants need to monitor the other three market sectors-the dollar, bonds,  
commodities and others.  The fact that the equity collapse was global in scope, and not limited
to the U.S. markets, many would seem to argue against such a narrow view and finally supplied enough arguments for the global analysis of the considered markets. Therefore, Murphy has focused on the
commodity, bond, stock and currency markets, globally. Among the many conclusions, He  presents many arguments  that the U.S. dollar contributed to the weakness in equities. 
 Moreover, he concludes that among the four considered sectors the role of the U.S. dollar is probably the least precise and the one most difficult to pin down \citep{bib:murp}. However, all Murphy's considerations are done on the basis of binary relations resulting from the binary correlations. In this paper we are going to extend his Intermarket Technical Analysis onto a hierarchy of relations including   two, three, four,.....$m$ -ary relations, where $m$ is number of sectors constituting the considered complex system.  In order to approach this idea  we apply  concept of  frustrations to the U.S. Intermarket which is constituted by the following sectors: Stock, Bond, Commodity, Currency and Gold. Therefore, for the further considerations we select the following list of indexes:\\
 S$\&$P500 - SPX, 
Treasury Bonds Prices - USB,
Commodity Research Bureau Futures Price Index - CRB,
U. S. Dollar Index - USD and  Gold  Index - XAU. \\
A distribution of entanglement signs between spins play the crucial role in typical models of spin glass. In the case of Intermarket the entanglement between sectors 
is described by the linear correlation coefficient. Therefore, the frustration in this system is determined by the distribution of correlation coefficient's signs Illustrations of the all concepts derived here are presented on the example data 1987/07/01-1987/12/31. Whereas,
in order to investigate dynamics of the considered hierarchy of frustrations we analyse  1983 - 2012  statistical data \citep{bib:NICK}. For a test we applied some data from \citep{bib:MURsite}. The correlation coefficients are calculated for each half of the year. The example correlation coefficients are presented in TABLE \ref{Table0}.
 \begin{table}[ht]
\renewcommand{\arraystretch}{1.3}
\caption{Correlation coefficients of the  sectors' indexes for the periods 1987/01/02-1987/06/30 and 1987/07/01-1987/12/31, above and below diagonal, respectively.}
\label{Table0}
\begin{center}
\begin{tabular}{|c||c|c|c|c|c|}
\hline
 &	CRB &	USB &	SPX &	USD &	XAU\\ \hline\hline
CRB	&1&	-0,115&	-0,003&	-0,380&	0,401\\ \hline
USB	&-0,144	&1	&0,544&	-0,271&	-0,666\\ \hline
SPX	&0,376	&0,617	&1	&-0,124&	-0,182 \\ \hline
USD	&0,129	&-0,085	&0,456	&1	&-0,195\\ \hline
XAU	&0,750	&-0,081	&0,235	&-0,351	&1\\ \hline
 \end{tabular}
\end{center}
\end{table}

\section{Frustrations in correlated subsystems}\label{defin}
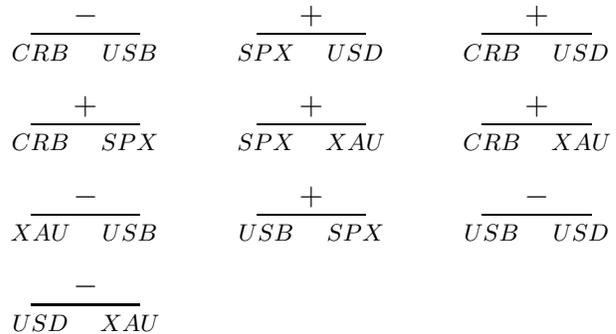
\begin{figure}
\begin{center}
\setlength{\unitlength}{0.12mm}
\begin{picture}(250,450)(0,-220)
\put(-203,150){\line(1,0){120}}
\put(-193,130){\small{\makebox(0,0){${ CRB}$}}}
\put(-94,130){\small{\makebox(0,0){$USB$}}}
\put(-143,170){\large{\makebox(0,0){${-}$}}}


\put(47,150){\line(1,0){120}}
\put(57,130){\small{\makebox(0,0){${ SPX}$}}}
\put(156,130){\small{\makebox(0,0){$USD$}}}
\put(107,170){\large{\makebox(0,0){${+}$}}}


\put(297,150){\line(1,0){120}}
\put(307,130){\small{\makebox(0,0){${ CRB}$}}}
\put(406,130){\small{\makebox(0,0){$USD$}}}
\put(357,170){\large{\makebox(0,0){${+}$}}}

\put(-203,50){\line(1,0){120}}
\put(-193,30){\small{\makebox(0,0){${ CRB}$}}}
\put(-94,30){\small{\makebox(0,0){$SPX$}}}
\put(-143,70){\large{\makebox(0,0){${+}$}}}


\put(47,50){\line(1,0){120}}
\put(57,30){\small{\makebox(0,0){${ SPX}$}}}
\put(156,30){\small{\makebox(0,0){$XAU$}}}
\put(107,70){\large{\makebox(0,0){${+}$}}}


\put(297,50){\line(1,0){120}}
\put(307,30){\small{\makebox(0,0){${ CRB}$}}}
\put(406,30){\small{\makebox(0,0){$XAU$}}}
\put(357,70){\large{\makebox(0,0){${+}$}}}

\put(-203,-50){\line(1,0){120}}
\put(-193,-70){\small{\makebox(0,0){${XAU}$}}}
\put(-94,-70){\small{\makebox(0,0){$USB$}}}
\put(-143,-30){\large{\makebox(0,0){${-}$}}}


\put(47,-50){\line(1,0){120}}
\put(57,-70){\small{\makebox(0,0){${USB}$}}}
\put(156,-70){\small{\makebox(0,0){$SPX$}}}
\put(107,-30){\large{\makebox(0,0){${+}$}}}


\put(297,-50){\line(1,0){120}}
\put(307,-70){\small{\makebox(0,0){${USB}$}}}
\put(406,-70){\small{\makebox(0,0){$USD$}}}
\put(357,-30){\large{\makebox(0,0){${-}$}}}


\put(-203,-150){\line(1,0){120}}
\put(-193,-170){\small{\makebox(0,0){${ USD}$}}}
\put(-94,-170){\small{\makebox(0,0){$XAU$}}}
\put(-143,-130){\large{\makebox(0,0){${-}$}}}



\end{picture}
\caption{\label{fig01}
 Graphic representation of the relation ${\bf {\hat{R}_{2}}}$ fort the period 1987/07/01-1987/12/31 reduced to (\ref{rho}) 
}
\end{center}
\end{figure}

Let us consider the following Intermarket's sectors represented by their indexes:
\begin{equation}
 S=\{SPX,USB,CRB,USD,XAU\}
\end{equation}
 and define the following binary relation:
\begin{equation}
\label{R}
{\bf {{\hat{R}}_{2}}}\subset S\times S\times\{-,+\},
\end{equation}
where $\{-,+\}$ is set of two labels corresponding to signs of correlation coefficients of the pairs belonging to $S\times S$.\\
 Let us determine  relation ${\bf {\hat{R}_{2}}}$ on  $S^{2}\times\{-,+\}$.  For the binary relation we use the following notation:  
\begin{equation}
\label{pm} 
(X,Y,\pm)\in{\bf {\hat{R}_{2}}}\equiv X{\bf {\hat{R}_{2}}}Y=\pm.
\end{equation}
Let  $\pm$ are determined by the signs of  Pearson's correlation coefficients  
\begin{equation}
\label{rho}
X{\bf \hat R_{2}}Y=sign(\rho(X,Y)),
\end{equation}
where  $X,Y\in S$  and 
\begin{equation}
\rho(X,Y)=Cov_{XY}/\sqrt{Var_{X}Var_{Y}} .
\end{equation}
Therefore the considered relation becomes to the following function (for the period 1987/07/01-1987/12/31):
\begin{eqnarray}
CRB{{\bf {\hat{R}_{2}}}}SPX=-,\hspace{2mm}CRB{{\bf {\hat{R}_{2}}}}XAU=+,\nonumber\\
 SPX{{\bf {\hat{R}_{2}}}}USB=-,\hspace{2mm}CRB{{\bf {\hat{R}_{2}}}}USD=-,\nonumber\\
CRB{{\bf {\hat{R}_{2}}}}USB=-,\hspace{2mm}USD{{\bf {\hat{R}_{2}}}}USB=+,\nonumber\\
\end{eqnarray}
The remaining mappings of this function are presented in Fig. \ref{fig01} (for the period 1987/07/01-1987/12/31).

The values of  the correlation coefficients concerning the selected sectors are presented in TABLE \ref{Table0}. 
Let us investigate the transitivity of ${\bf {\hat{R}_{2}}}$. Therefore we have to determine superposition of the two relations like the following example: $CRB{\bf {\hat{R}_{2}}}SPX{\bf {\hat{R}_{2}}}USB$. Let $X,Y,Z$  
are different members of $S$. According to least squares estimates \citep{bib:Brandt} we can write down the following linear approximations:
\begin{eqnarray}
Y=a_{XY}\cdot X+b_{XY},\label{XY}\\
Z=a_{YZ}\cdot Y+b_{YZ}\label{YZ}.
\end{eqnarray}
Let us create a superposition of  (\ref{XY}) and (\ref{YZ}):
\begin{equation}
\label{Z}
Z=a_{XZ}\cdot X+b_{XZ},
\end{equation}
  where 
\begin{eqnarray}
a_{XZ}=a_{XY}\cdot  a_{YZ},\label{ab1}\\
b_{XZ}=b_{XY}\cdot a_{YZ}+b_{YZ}.\label{ab2}
\end{eqnarray}
According to the least squares estimates $sign(a_{XY})=sign(\rho(X,Y))=\Phi_{R_{2}}(X,Y)$, therefore (\ref{rho}) takes the following form:
\begin{equation}
\label{rho_a}
X{\bf \hat R_{2}}Y=\Phi_{R_{2}}(X,Y).
\end{equation}
Combining (\ref{rho})-(\ref{rho_a}) we derive the following superposition's rule for ${\bf {\hat{R}_{2}}}$:
\begin{equation}
\label{RR_aa}
X{\bf {\hat{R}_{2}}}Y{\bf {\hat{R}_{2}}}Z=\Phi_{R_{2}}(X,Y)\,\Phi_{R_{2}}(Y,Z).
\end{equation}
Therefore we formulate and prove the following theorem:\\
{\em Theorem 1}\\
Let $\forall  X,Y,Z\in S_{T}\subset S$  
\begin{eqnarray} 
X{\bf \hat R_{2}}Y=\Phi_{R_{2}}(X,Y)\land Y{\bf \hat R_{2}}Z=\Phi_{R_{2}}(Y,Z)\land \nonumber\\
X{\bf \hat R_{2}}Z=\Phi_{R_{2}}(X,Z), \label{tran1}
\end{eqnarray}
then $\bf{\hat{R}_{2}}$ is transitive in $S_{T}$ if       
\begin{equation}
\label{XYYZZX}
\Phi_{R_{2}}(X,Y)\,\Phi_{R_{2}}(Y,Z)\,\Phi_{R_{2}}(X,Z)=+.
\end{equation}
{\em Proof}\\
By the definition ${\bf {\hat{R}_{2}}}$ is transitive if $X{\bf \hat R_{2}}Y=\Phi_{R_{2}}(X,Y)\land Y{\bf \hat R_{2}}Z=\Phi_{R_{2}}(Y,Z)\Rightarrow X{\bf \hat R_{2}}Z=\Phi_{R_{2}}(X,Z)$.  
On the basis of this definition and (\ref{RR_aa}) as well as (\ref{tran1}) we derive that ${\bf {\hat{R}_{2}}}$ is transitive if:
\begin{equation}
\label{XYYZ}
\Phi_{R_{2}}(X,Y)\,\Phi_{R_{2}}(Y,Z)=\Phi_{R_{2}}(X,Z).
\end{equation}
Multiplying (\ref{XYYZ}) by $\Phi_{R_{2}}(X,Z)$ we derive (\ref{XYYZZX}). $\blacksquare$

Taking into account above theorem we obtain:
\begin{eqnarray}
CRB{\bf {\hat{R}_{2}}}USD{\bf {\hat{R}_{2}}}USB= CRB{\bf {\hat{R}_{2}}}USB,\nonumber\\
CRB{\bf {\hat{R}_{2}}}USB{\bf {\hat{R}_{2}}}XAU= CRB{\bf {\hat{R}_{2}}}XAU,\label{notran}\\
USD{\bf {\hat{R}_{2}}}SPX{\bf {\hat{R}_{2}}}XAU= USD{\bf {\hat{R}_{2}}}XAU.\nonumber\\
\cdots\hspace{15mm}\cdots\hspace{15mm}\cdots\hspace{15mm}\cdots\nonumber
\end{eqnarray}

\begin{eqnarray}
CRB{\bf {\hat{R}_{2}}}SPX{\bf {\hat{R}_{2}}}USB\neq CRB{\bf {\hat{R}_{2}}}USB,\nonumber\\
CRB{\bf {\hat{R}_{2}}}USD{\bf {\hat{R}_{2}}}XAU\neq CRB{\bf {\hat{R}_{2}}}XAU,\label{tran}\\
USB{\bf {\hat{R}_{2}}}USD{\bf {\hat{R}_{2}}}XAU\neq USB{\bf {\hat{R}_{2}}}XAU,\nonumber\\
\cdots\hspace{15mm}\cdots\hspace{15mm}\cdots\hspace{15mm}\cdots\nonumber
\end{eqnarray}

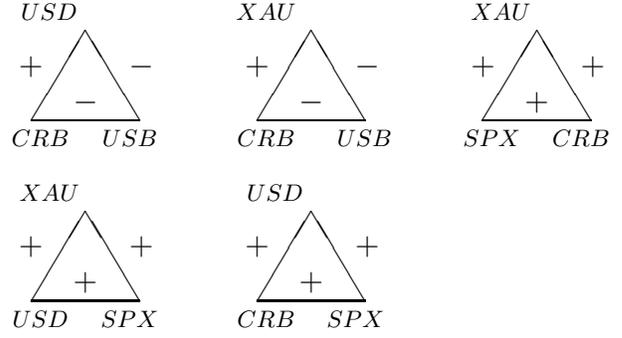
\begin{figure}
\begin{center}
\setlength{\unitlength}{0.12mm}
\begin{picture}(250,450)(0,-220)
\put(-203,50){\line(3,5){60}}
\put(-83,50){\line(-3,5){60}}
\put(-203,50){\line(1,0){120}}
\put(-193,30){\small{\makebox(0,0){${ CRB}$}}}
\put(-183,170.2){\small {\makebox(0,0){${USD}$}}}
\put(-94,30){\small{\makebox(0,0){$USB$}}}
\put(-143,70){\large{\makebox(0,0){${-}$}}}
\put(-81,110.1){\large{\makebox(0,0){${-}$}}}
\put(-203.40,110.1){\large{\makebox(0,0){${+}$}}}
\put(47,50){\line(3,5){60}}
\put(167,50){\line(-3,5){60}}
\put(47,50){\line(1,0){120}}
\put(57,30){\small{\makebox(0,0){${ CRB}$}}}
\put(57,170.2){\small {\makebox(0,0){${XAU}$}}}
\put(166,30){\small{\makebox(0,0){$USB$}}}
\put(107,70){\large{\makebox(0,0){${-}$}}}
\put(169.1,110.1){\large{\makebox(0,0){${-}$}}}
\put(47.40,110.1){\large{\makebox(0,0){${+}$}}}
\put(297,50){\line(3,5){60}}
\put(417,50){\line(-3,5){60}}
\put(297,50){\line(1,0){120}}
\put(307,30){\small{\makebox(0,0){${ SPX}$}}}
\put(317,170.2){\small {\makebox(0,0){${XAU}$}}}
\put(406,30){\small{\makebox(0,0){$CRB$}}}
\put(357,70){\large{\makebox(0,0){${+}$}}}
\put(419.1,110.1){\large{\makebox(0,0){${+}$}}}
\put(297.40,110.1){\large{\makebox(0,0){${+}$}}}

\put(-203,-150){\line(3,5){60}}
\put(-83,-150){\line(-3,5){60}}
\put(-203,-150){\line(1,0){120}}
\put(-193,-170){\small{\makebox(0,0){${ USD}$}}}
\put(-183,-30){\small {\makebox(0,0){${XAU}$}}}
\put(-94,-170){\small{\makebox(0,0){$SPX$}}}
\put(-143,-130){\large{\makebox(0,0){${+}$}}}
\put(-81,-90.1){\large{\makebox(0,0){${+}$}}}
\put(-203.40,-90.1){\large{\makebox(0,0){${+}$}}}
\put(47,-150){\line(3,5){60}}
\put(167,-150){\line(-3,5){60}}
\put(47,-150){\line(1,0){120}}
\put(57,-170){\small{\makebox(0,0){${ CRB}$}}}
\put(67,-30.2){\small {\makebox(0,0){${USD}$}}}
\put(156,-170){\small{\makebox(0,0){$SPX$}}}
\put(107,-130){\large{\makebox(0,0){${+}$}}}
\put(169.1,-90){\large{\makebox(0,0){${+}$}}}
\put(47.40,-90.1){\large{\makebox(0,0){${+}$}}}

\end{picture}
\caption{
\label{fig1}
{ The subspace $V_{T}$ for the  period.1987/07/01-1987/12/31}
}
\end{center}
\end{figure}

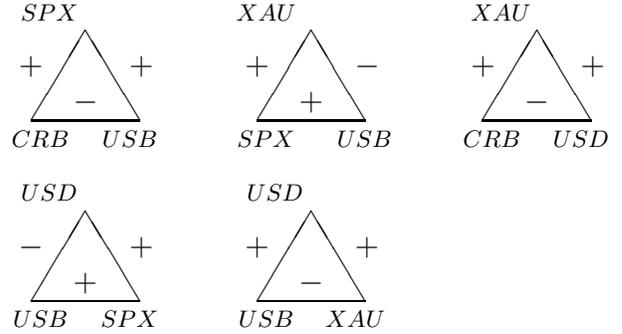
\begin{figure}
\begin{center}
\setlength{\unitlength}{0.12mm}
\begin{picture}(250,450)(0,-220)
\put(-203,50){\line(3,5){60}}
\put(-83,50){\line(-3,5){60}}
\put(-203,50){\line(1,0){120}}
\put(-193,30){\small{\makebox(0,0){${ CRB}$}}}
\put(-183,170.2){\small {\makebox(0,0){${SPX}$}}}
\put(-94,30){\small{\makebox(0,0){$USB$}}}
\put(-143,70){\large{\makebox(0,0){${-}$}}}
\put(-81,110.1){\large{\makebox(0,0){${+}$}}}
\put(-203.40,110.1){\large{\makebox(0,0){${+}$}}}
\put(47,50){\line(3,5){60}}
\put(167,50){\line(-3,5){60}}
\put(47,50){\line(1,0){120}}
\put(57,30){\small{\makebox(0,0){${ SPX}$}}}
\put(57,170.2){\small {\makebox(0,0){${XAU}$}}}
\put(166,30){\small{\makebox(0,0){$USB$}}}
\put(107,70){\large{\makebox(0,0){${+}$}}}
\put(169.1,110.1){\large{\makebox(0,0){${-}$}}}
\put(47.40,110.1){\large{\makebox(0,0){${+}$}}}

\put(297,50){\line(3,5){60}}
\put(417,50){\line(-3,5){60}}
\put(297,50){\line(1,0){120}}
\put(307,30){\small{\makebox(0,0){${ CRB}$}}}
\put(317,170.2){\small {\makebox(0,0){${XAU}$}}}
\put(406,30){\small{\makebox(0,0){$USD$}}}
\put(357,70){\large{\makebox(0,0){${-}$}}}
\put(419.1,110.1){\large{\makebox(0,0){${+}$}}}
\put(297.40,110.1){\large{\makebox(0,0){${+}$}}}
\put(-203,-150){\line(3,5){60}}
\put(-83,-150){\line(-3,5){60}}
\put(-203,-150){\line(1,0){120}}
\put(-193,-170){\small{\makebox(0,0){${USB}$}}}
\put(-183,-30){\small {\makebox(0,0){${USD}$}}}
\put(-94,-170){\small{\makebox(0,0){$SPX$}}}
\put(-143,-130){\large{\makebox(0,0){${+}$}}}
\put(-81,-90.1){\large{\makebox(0,0){${+}$}}}
\put(-203.40,-90.1){\large{\makebox(0,0){${-}$}}}

\put(47,-150){\line(3,5){60}}
\put(167,-150){\line(-3,5){60}}
\put(47,-150){\line(1,0){120}}
\put(57,-170){\small{\makebox(0,0){${ USB}$}}}
\put(67,-30.2){\small {\makebox(0,0){${USD}$}}}
\put(156,-170){\small{\makebox(0,0){$XAU$}}}
\put(107,-130){\large{\makebox(0,0){${-}$}}}
\put(169.1,-90){\large{\makebox(0,0){${+}$}}}
\put(47.40,-90.1){\large{\makebox(0,0){${+}$}}}


\end{picture}
\caption{
\label{fig2}
{ The subspace $V_{F}$ for the period 1987/07/01-1987/12/31. (The signs of the correlations are supplied for  convenience in testing of ${\bf {\hat{R}_{2}}}$ with respect to the transitivity) }
}
\end{center}
\end{figure}

The remaining tests of transitivity are presented in Fig. \ref{fig1} and Fig. \ref{fig2}. \\
Let  us call the following subset $\{X\neq Y\neq Z\neq X\}\subset S$ a plaquette, let $V$ be set of the all plaquettes created in $S$. 
We can see that the transitivity decomposes $V$   
into two components: $V={V_{T}\cup V_{F}}$, 
where $\forall\{ X,Y,Z\}\in V_{T}$ the relation ${\bf {\hat{R}_{2}}}$ 
is transitive and it is not transitive $\forall \{X,Y,Z\}\in V_{F}$, 
(N, F - mean no frustration and frustration, respectively).

{\em Definition} \\Lacking of ${\bf {\hat{R}_{2}}}$'s transitivity  in $\{X,Y,Z\}\in V_{F}$ we call  frustration of $\{X,Y,Z\}$  with respect to ${\bf {\hat{R}_{2}}}$. \\

\subsection{Hierarchy of frustrations in correlated subsystems}
Let us consider the following Intermarket's sectors represented by their indexes: $S=\{SPX,USB,CRB,USD,XAU\}$ and define the following hierarchy of 
relations:
\begin{equation}
\label{R}
{\bf {\hat{R}_{2}}}\subset S\times S\times\{-,+\},
\end{equation}
where $\{-,+\}$ is set of two labels corresponding to signs of correlation coefficients of the pairs belonging to $S\times S$.
\begin{equation}
\label{Q}
{\bf {\hat{R}_{3}}}\subset S\times S\times S\times\{F_{R_{2}},T_{R_{2}}\},
\end{equation}
where $\{F_{R_{2}},T_{R_{2}}\}$ is set of two labels: $F_{R_{2}}$ - frustration, $T_{R_{2}}$ - no frustration with respect to transitivity of ${\bf {\hat{R}_{2}}}$. 
\begin{equation}
\label{P}
{\bf {\hat{R}_{4}}}\subset S\times S\times S\times S\times\{F_{R_{3}},T_{R_{3}}\},
\end{equation}
where $\{F_{R_{3}},T_{R_{3}}\}$ is set of two labels: $F_{R_{3}}$ - frustration, $T_{R_{3}}$ - no frustration with respect to transitivity of ${\bf {\hat{R}_{3}}}$.\\
\begin{equation}
\label{O}
{\bf {\hat{R}_{5}}}\subset S\times S\times S\times S\times S \times \{F_{R_{4}},T_{R_{4}}\},
\end{equation}
where $\{F_{R_{4}},T_{R_{4}}\}$ is set of two labels: $F_{R_{4}}$ - frustration, $T_{R_{4}}$ - no frustration with respect to transitivity of ${\bf {\hat{R}_{4}}}$.\\
\subsubsection{The ternary relation}\label{II.1.1}
\begin{figure}
\begin{center}
\setlength{\unitlength}{0.12mm}
\begin{picture}(250,450)(0,-220)
\put(-203,50){\line(3,5){60}}
\put(-83,50){\line(-3,5){60}}
\put(-203,50){\line(1,0){120}}
\put(-193,30){\small{\makebox(0,0){${ CRB}$}}}
\put(-183,170.2){\small {\makebox(0,0){${SPX}$}}}
\put(-94,30){\small{\makebox(0,0){$USB$}}}
\put(-143,95){\large{\makebox(0,0){${{\textbf{+}}}$}}}

\put(47,50){\line(3,5){60}}
\put(167,50){\line(-3,5){60}}
\put(47,50){\line(1,0){120}}
\put(57,30){\small{\makebox(0,0){${ SPX}$}}}
\put(57,170.2){\small {\makebox(0,0){${XAU}$}}}
\put(166,30){\small{\makebox(0,0){$USD$}}}
\put(107,95){\large{\makebox(0,0){${\textbf{+}}$}}}

\put(297,50){\line(3,5){60}}
\put(417,50){\line(-3,5){60}}
\put(297,50){\line(1,0){120}}
\put(307,30){\small{\makebox(0,0){${ CRB}$}}}
\put(317,170.2){\small {\makebox(0,0){${XAU}$}}}
\put(406,30){\small{\makebox(0,0){$USD$}}}
\put(357,95){\large{\makebox(0,0){${\textbf{+}}$}}}


\put(-203,-150){\line(3,5){60}}
\put(-83,-150){\line(-3,5){60}}
\put(-203,-150){\line(1,0){120}}
\put(-193,-170){\small{\makebox(0,0){${ CRB}$}}}
\put(-183,-30){\small {\makebox(0,0){${XAU}$}}}
\put(-94,-170){\small{\makebox(0,0){$SPX$}}}
\put(-143,-105){\large{\makebox(0,0){${\textbf{+}}$}}}

\put(47,-150){\line(3,5){60}}
\put(167,-150){\line(-3,5){60}}
\put(47,-150){\line(1,0){120}}
\put(57,-170){\small{\makebox(0,0){${ CRB}$}}}
\put(67,-30.2){\small {\makebox(0,0){${USD}$}}}
\put(156,-170){\small{\makebox(0,0){$SPX$}}}
\put(107,-105){\large{\makebox(0,0){${\textbf{+}}$}}}

\put(297,-150){\line(3,5){60}}
\put(417,-150){\line(-3,5){60}}
\put(297,-150){\line(1,0){120}}
\put(307,-170){\small{\makebox(0,0){${USB}$}}}
\put(317,-30.2){\small {\makebox(0,0){${XAU}$}}}
\put(406,-170){\small{\makebox(0,0){$USD$}}}
\put(357,-105){\large{\makebox(0,0){${\textbf{+}}$}}}


\end{picture}
\caption{
\label{fig02}
{ Subrelation $XYZ{\bf {\hat{R}_{3T}}}={\textbf{+}}$} }
\end{center}
\end{figure}
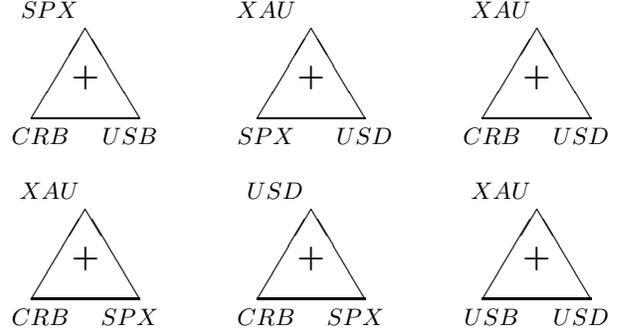

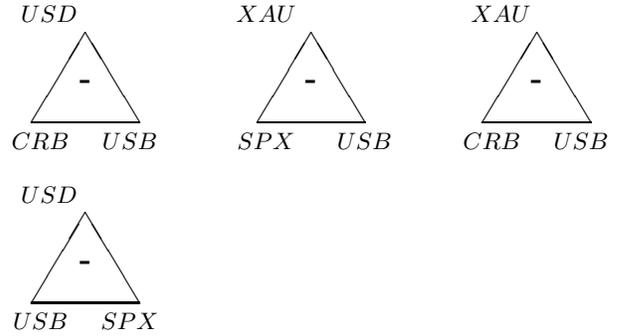
\begin{figure}
\begin{center}
\setlength{\unitlength}{0.12mm}
\begin{picture}(250,450)(0,-220)
\put(-203,50){\line(3,5){60}}
\put(-83,50){\line(-3,5){60}}
\put(-203,50){\line(1,0){120}}
\put(-193,30){\small{\makebox(0,0){${ CRB}$}}}
\put(-183,170.2){\small {\makebox(0,0){${USD}$}}}
\put(-94,30){\small{\makebox(0,0){$USB$}}}
\put(-143,95){\large{\makebox(0,0){${\textbf{-}}$}}}

\put(47,50){\line(3,5){60}}
\put(167,50){\line(-3,5){60}}
\put(47,50){\line(1,0){120}}
\put(57,30){\small{\makebox(0,0){${ SPX}$}}}
\put(57,170.2){\small {\makebox(0,0){${XAU}$}}}
\put(166,30){\small{\makebox(0,0){$USB$}}}
\put(107,95){\large{\makebox(0,0){${\textbf{-}}$}}}

\put(297,50){\line(3,5){60}}
\put(417,50){\line(-3,5){60}}
\put(297,50){\line(1,0){120}}
\put(307,30){\small{\makebox(0,0){${ CRB}$}}}
\put(317,170.2){\small {\makebox(0,0){${XAU}$}}}
\put(406,30){\small{\makebox(0,0){$USB$}}}
\put(357,95){\large{\makebox(0,0){${\textbf{-}}$}}}

\put(-203,-150){\line(3,5){60}}
\put(-83,-150){\line(-3,5){60}}
\put(-203,-150){\line(1,0){120}}
\put(-193,-170){\small{\makebox(0,0){${USB}$}}}
\put(-183,-30){\small {\makebox(0,0){${USD}$}}}
\put(-94,-170){\small{\makebox(0,0){$SPX$}}}
\put(-143,-105){\large{\makebox(0,0){${\textbf{-}}$}}}


\end{picture}
\caption{
\label{fig03}
{  Subrelation $XYZ{\bf {\hat{R}_{3F}}}={\textbf{-}}$}
}
\end{center}
\end{figure}

Now we are ready to define ${\bf {\hat{R}_{3}}}$. \\
{\em Definition} Let us create all possible points $(X,Y,Z)\in S^{3}$, 
where $X,Y,Z\in S$ and $X\neq Y, Y\neq Z, Z\neq X $ 
and define  ${\bf {\hat{R}_{3}}}$ in the following way: 
if frustration with respect to transitivity in $\bf {\hat{R}_{2}}$ 
occurs in ${X,Y,Z}$ then the point $(X,Y,Z)$ is mapped to $F_{R_{2}}$
and $(X,Y,Z,F_{R_{2}}) \in {\bf {\hat{R}_{3F}}}$ else $(X,Y,Z)$ 
is mapped to $T_{R_{2}}$ and $(X,Y,Z,T_{R_{2}})\in {\bf {\hat{R}_{3T}}}$. 
Finally 
${\bf {\hat{R}_{3}}}={\bf {\hat{R}_{3F}}} \cup {\bf {\hat{R}_{3T}}}$. 
The subsets corresponding to ${T_{R_{2}}}$ and ${F_{R_{2}}}$:
\begin{eqnarray}
{\bf {\hat{R}_{3T}}}\subset S\times S\times S\times\{T_{R_{2}}\},\label{QN}\\
{\bf {\hat{R}_{3F}}}\subset S\times S\times S\times\{F_{R_{2}}\},\label{QF}
\end{eqnarray}
are presented in Fig. \ref{fig02} and Fig. \ref{fig03}, respectively.
It occurs the following relation between ${\bf {\hat{R}_{3}}}$ and ${\bf {\hat{R}_{2}}}$:
\begin{equation}
\label{XYZRxy}
XYZ{\bf {\hat{R}_{3}}}=X{\bf {\hat{R}_{2}}}Y\land Y{\bf {\hat{R}_{2}}}Z\land Z{\bf {\hat{R}_{2}}}X.
\end{equation}
Therefore, similarly to (\ref{rho_a})  and according to (\ref{XYZRxy})  
we determine  the values of  ${\bf {\hat{R}_{3}}}$ in the following way: 
\begin{eqnarray}
{\it \Phi_{R_{3}}(X,Y,Z)}=\Phi_{R_{2}}(X,Y)\,\Phi_{R_{2}}({Y,Z})\,\Phi_{R_{2}}({X,Z}),\nonumber\\
{\it \Phi_{R_{3}}(T,X,Z)}=\Phi_{R_{2}}({T,X})\,\Phi_{R_{2}}({X,Z})\,\Phi_{R_{2}}({T,Z}),\label{QXYZ}\\
{\it \Phi_{R_{3}}(T,X,Y)}=\Phi_{R_{2}}({T,X})\,\Phi_{R_{2}}(X,Y)\,\Phi_{R_{2}}({T,Y}),\nonumber\\
{\it \Phi_{R_{3}}(T,Y,Z)}=\Phi_{R_{2}}({T,Y})\,\Phi_{R_{2}}({Y,Z})\,\Phi_{R_{2}}({T,Z}).\nonumber
\end{eqnarray}

In order to extent notion of frustration into ${\bf {\hat{R}_{3}}}$ 
we have to define the transitivity with respect to this relation.\\
{\em Definition}\\ Let 
\begin{equation}
\label{tranS3}
\{(X,Y,Z),(T,X,Z),(T,X,Y),(T,Y,Z)\}\subset S^{3}.
\end{equation}
 If  
\begin{eqnarray}
XYZ{\bf {\hat{R}_{3}}}=\Phi_{R_{3}}(X,Y,Z) \land \nonumber\\
XTZ{\bf {\hat{R}_{3}}}=\Phi_{R_{3}}(X,T,Z)\land\label{XYZQ}\\
XTY{\bf {\hat{R}_{3}}}=\Phi_{R_{3}}(X,T,Y)\Rightarrow \nonumber\\
TYZ{\bf {\hat{R}_{3}}}=\Phi_{R_{3}}(T,Y,Z),\nonumber
\end{eqnarray}
then  $\bf {\hat{R}_{3}}$ is transitive in (\ref{tranS3}). \\
{\em Theorem 2}\\
Let $\bf {\hat R_{3}}$  be transitive in \\$\{(X,Y,Z),(T,X,Z),(T,X,Y),(T,Y,Z)\}$:
\begin{eqnarray} 
XYZ{\bf \hat R_{3}}=\Phi_{R_{3}}(X,Y,Z)\land \nonumber\\
TXZ{\bf \hat R_{3}}=\Phi_{R_{3}}(T,X,Z)\land \nonumber\\
TXY{\bf \hat R_{3}}=\Phi_{R_{3}}(X,T,Y)\land \nonumber\\
TYZ{\bf \hat R_{3}}=\Phi_{R_{3}}(T,Y,Z), \label{tran31}
\end{eqnarray}
then 
\begin{eqnarray}
\Phi_{R_{3}}(X,Y,Z)\,\Phi_{R_{3}}(X,T,Z)\nonumber\\
\Phi_{R_{3}}(X,T,Y)\,\Phi_{R_{3}}(T,Y,Z)=+.\label{3XYYZZX}
\end{eqnarray}
{\em Proof}\\
By the definition ${\bf {\hat{R}_{3}}}$ is transitive if $XYZ{\bf \hat R_{3}}=\Phi_{R_{3}}(X,Y,Z)\land TYZ{\bf \hat R_{3}}=\Phi_{R_{3}}(T,Y,Z)\land TXZ{\bf \hat R_{3}}=\Phi_{R_{3}}(T,X,Z)\Rightarrow TXY{\bf \hat R_{3}}=\Phi_{R_{3}}(T,X,Y)$.  
Taking into account (\ref{tranS3}) and  $\Phi_{R_{2}}^{2}=+1$ as well as  (\ref{QXYZ}) we derive that ${\bf {\hat{R}_{3}}}$ is transitive if:
\begin{eqnarray}
{\it \Phi_{R_{3}}(T,X,Z)}=\label{Qcomp}\\
{\it \Phi_{R_{3}}(T,X,Y)}\,{\it \Phi_{R_{3}}(X,Y,Z)}\,{\it \Phi_{R_{3}}(T,Y,Z)}.\nonumber
\end{eqnarray}
Multiplying (\ref{Qcomp}) by ${\it \Phi_{R_{3}}(T,X,Z)}$ we get the thesis.
$\blacksquare$
\subsubsection{The 4-ary and 5-ary relations}
Extending results of $\ref{II.1.1}$ into the 4-ary and 5-ary relations (\ref{P}) and (\ref{O}), respectively we define the transitivity and frustration for the 4 and 5 point complexes which are presented in Figure \ref{fig24} and Figure \ref{fig25}, respectively. For the considered $S$ system there are five 4-ary relations and one 5-ary relation. 
Extending (\ref{QXYZ}) on $\bf {\hat{R}_{4}}$ we derive the following relation:
\begin{equation}
\label{PHI_R4}
\Phi_{R_{4}}(X_{1},X_{2},X_{3},X_{4})= \Phi_{R_{3}}(X_{1},X_{2},X_{3})\prod_{i=1}^{3}\Phi_{R_{2}}(X_{i},X_{4})\end{equation}
For presentation of transitivity and frustration in  
${\bf {\hat{R_{4}}}}$ we calculate $\Phi_{R_{4}}$ for the both 
selected relations (Figure \ref{fig24}): 
$\Phi_{R_{4}}(CRB,SPX,USB,USD)=+1, \Phi_{R_{4}}(XAU,SPX,USB,USD)=-1$.  
Extending (\ref{PHI_R4}) on ${\bf {\hat{R}_{5}}}$ we derive the following 
value of the transitivity for the whole considered system: 
$\Phi_{R_{5}}(S)=+1$. Therefore, ${\bf \hat{R_{4}}} $ is not transitive 
in $S-CRB$, whereas it is transitive in $S-XAU$ as well. Since  
${\bf {\hat{R}_{5}}}$ is transitive in $S$ whereas it is not transitive 
in $S-CRB$ we derive the following conclusion: 
in the period 1987/07/01-1987/12/31 $CRB$ has been played an ordering role in the considered Intermarket.  Thus, e.g. relating the transitivity's measures of  $\Phi_{R_{4}}$ and $\Phi_{R_{5}}$ we investigate roles of the all Intermarket's sectors during 1983-2012 (Section \ref{anal}).
\begin{figure}
\begin{center}
\setlength{\unitlength}{0.18mm}
\begin{picture}(130,170)(-90,-10)
\put(-203,50){\line(3,5){60}}
\put(-83,50){\line(-3,5){60}}
\put(-203,50){\line(1,0){120}}
\put(-143,50){\line(-3,5){30}}
\put(-143,50){\line(3,5){30}}
\put(-173,100){\line(1,0){60}}
\put(-210,30){\small{\makebox(0,0){${CRB}$}}}
\put(-143,170.2){\small {\makebox(0,0){${CRB}$}}}
\put(-74,30){\small{\makebox(0,0){$CRB$}}}
\put(-143,30){\small{\makebox(0,0){${SPX}$}}}
\put(-81,100.1){\small{\makebox(0,0){${USB}$}}}
\put(-203.40,100.1){\small{\makebox(0,0){${USD}$}}}
\put(-177,65){\large{\makebox(0,0){${+}$}}}
\put(-108,65){\large{\makebox(0,0){${+}$}}}
\put(-143,80){\large{\makebox(0,0){${-}$}}}
\put(-143,120){\large{\makebox(0,0){${-}$}}}
\put(-143,0){\large{\makebox(0,0){$\in V_{T}$}}}
\put(17,50){\line(3,5){60}}
\put(137,50){\line(-3,5){60}}
\put(17,50){\line(1,0){120}}
\put(77,50){\line(-3,5){30}}
\put(77,50){\line(3,5){30}}
\put(47,100){\line(1,0){60}}
\put(10,30){\small{\makebox(0,0){${XAU}$}}}
\put(77,170.2){\small {\makebox(0,0){${XAU}$}}}
\put(146,30){\small{\makebox(0,0){$XAU$}}}
\put(77,30){\small{\makebox(0,0){${SPX}$}}}
\put(144,100.1){\small{\makebox(0,0){${USB}$}}}
\put(17.40,100.1){\small{\makebox(0,0){${USD}$}}}
\put(43,65){\large{\makebox(0,0){${+}$}}}
\put(112,65){\large{\makebox(0,0){${-}$}}}
\put(77,80){\large{\makebox(0,0){${-}$}}}
\put(77,120){\large{\makebox(0,0){${+}$}}}
\put(77,0){\large{\makebox(0,0){$\in V_{F}$}}}
\end{picture}
\caption{\label{fig24}{ 
Subrelation ${\bf {\hat{R}_{4F}}}$.}}
\end{center}
\end{figure}
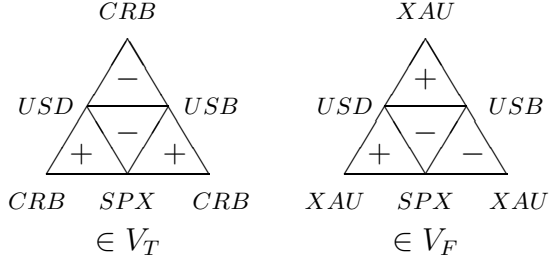

\begin{figure}
\begin{center}
\setlength{\unitlength}{0.15mm}
\begin{picture}(150,190)(-60,40)
\put(50,70){\line(1,1){50}}
\put(0,220){\line(-6,-1){138}}
\put(99,120){\line(1,-1){69.5}}
\put(50,70){\line(6,-1){118.5}}
\put(-100,120){\line(-1,2){38.5}}
\put(-100,120){\line(1,0){200}}
\put(-100,120){\line(3,-1){150}}
\put(-100,120){\line(1,1){100}}
\put(50,70){\line(-1,3){50}}
\put(100,120){\line(-1,1){100}}
\put(-132,120){\small{\makebox(0,0){${SPX}$}}}
\put(140,120){\small {\makebox(0,0){${XAU}$}}}
\put(48,50){\small{\makebox(0,0){$USD$}}}
\put(-10,233){\small{\makebox(0,0){$USB$}}}
\put(-170,195){\small{\makebox(0,0){$CRB$}}}
\put(199,55){\small{\makebox(0,0){$CRB$}}}
\end{picture}
\caption{
\label{fig25}{ Subrelation ${\bf {\hat{R}_{5F}}}$.}}
\end{center}
\end{figure}
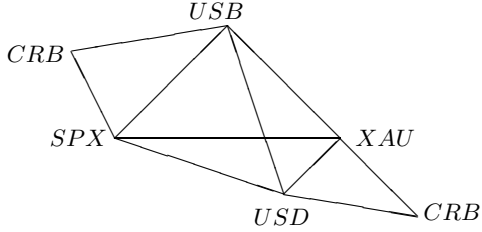
\subsubsection{The  n-ary relations}
Derivation of ${\bf \hat R_{3}},{\bf \hat R_{4}},{\bf \hat R_{5}},$ and their properties suggests the following algorithm for creation of the ${\bf \hat R_{n}}$ and investigation of the properties:\\
1. Write down the relation between ${\bf \hat R_{n}}$ and ${\bf \hat R_{n-1}}$. Let $(X_{1},X_{2}\dots X_{n})\in S^{n}$,
in  the following form:
\begin{equation}
\label{N_N_1}
X_{1}X_{2}\dots X_{n}{\bf {\hat{R}_{n}}}=\bigwedge_{i=1}^{ n }X_{1}X_{2}\dots X_{i-1}X_{i+1}\dots X_{n}{\bf \hat R_{n-1}},
\end{equation}
where $X_{n+1}=X_{1}$\\
2. Express $\Phi_{n}(X_{1},X_{2},\dots X_{n})$ by $\Phi_{2}(X_{i},X_{j})$, where $i,j=1,2\dots n$:
\begin{equation}
\label{N_2}
 \Phi_{n}(X_{1},X_{2},\dots X_{n})=\prod_{i<j\le n}\Phi_{R_{2}}(X_{i},X_{j}).
\end{equation}
3. Define transitivity of ${\bf {\hat{R}_{n}}}$.\\ Let $(X_{0},X_{1},X_{2},\dots X_{i-1},X_{i+1},\dots X_{n})\in S^{n}$, where $i=1,2,3\dots n$ and $X_{n+1}=X_{0}$\\ If the following relation occurs:
\begin{widetext}
\begin{equation}\label{tran_N}
\bigwedge_{i=1}^{n}\left(X_{0}X_{1}X_{2}\dots X_{i-1}X_{i+1}\dots X_{n}{\bf \hat R_{n}}=\Phi_{R_{n}}(X_{0},X_{1},X_{2},\dots X_{i-1},X_{i+1},\dots X_{n})\right)\\
 \rightarrow X_{1}X_{2}\dots X_{n}{\bf \hat R_{n}},
\end{equation}
\end{widetext}
then ${\bf {\hat{R}_{n}}}$ is transitive, else the subsystem $(X_{0},X_{1},X_{2},\dots X_{n})\in S^{n+1}$ is frustrated with respect to 
${\bf{\hat{R}_{n}}} $. \\
4. Derive recurrent formula for $\Phi_{R_{n}}(X_{1},X_{2},\dots X_{n})$:
\begin{widetext}
\begin{equation} 
\label{RECFi}
\Phi_{R_{n}}(X_{1},X_{2},\dots X_{n})=\Phi_{R_{n-1}}(X_{1},X_{2},\dots X_{n-1})\prod_{i=1}^{n-1}\Phi_{R_{2}}(X_{i},X_{n}).
\end{equation}
\end{widetext}
5. Derive the superposition rules for $\Phi_{R_{n}}(X_{1},X_{2},\dots X_{n})$. Writing down the complete system of (\ref{RECFi}) and performing elimination of the all $\Phi_{R_{2}}(X_{i},X_{j})$ correlation coefficients  we derive:
\begin{widetext}
\begin{equation}\label{supp_N}
\prod_{i=1}^{n}\Phi_{R_{n}}(X_{0},X_{1},X_{2},\dots X_{i-1},X_{i+1},\dots X_{n})=\Phi_{R_{n}}(X_{1}X_{2}\dots X_{n}).
\end{equation}
\end{widetext}

 \section{Measures of transitivites}\label{measure}
For each relation belonging to the hierarchy  
$\{ {\bf {{\hat{R}}_{2}}},{\bf {\hat{R}_{3}}},{\bf {\hat{R}_{4}}},
{\bf {\hat{R}_{5}}}, \dots \}$ the values of $\Phi_{R_{n}}=\pm 1$  
correspond to transitivity or frustration, respectively. 
However, they do not describe "how much" considered system 
is transitive or how much frustrated. In order to derive such 
a measure we come back to the definition of ${\bf {\hat{R}_{2}}}$. 
Let us note that $\rho(X,Y)=\Phi_{R_{2}}(X,Y)\cdot |(\rho(X,Y)) | 
=\rho_{R_{2}}(X,Y)$, where the first factor informs whether $X$ and $Y$ 
are correlated or anticorrelated, whereas the second one 
describes "how much". Therefore, we renamed  
$\rho(X,Y)$ into $\rho_{R_{2}}(X,Y)$ as an accepted measure of 
${\bf {\hat{R}_{2}}}$. Continuing this way and taking into account 
(\ref{QXYZ}) we derive the measures of 
${\bf {\hat{R}_{3}}}, {\bf {\hat{R}_{4}}}$ and ${\bf {\hat{R}_{5}}}$:
\begin{widetext}
\begin{eqnarray}
\rho_{R_{3}}(X,Y,Z)=\Phi_{R_{3}}(X,Y,Z)\cdot |\rho(X,Y) | \cdot\
 |\rho(Y,Z)| \cdot |\rho(X,Z) | =\rho(X,Y)  \cdot\
 \rho(Y,Z) \cdot \rho(X,Z)\label{MQ}.
\end{eqnarray}
Therefore,
\begin{eqnarray}
\rho_{R_{4}}(T,X,Y,Z)=\rho(X,Y) \cdot
\rho(Y,Z)\cdot \rho(X,Z) \cdot \rho(T,X)\cdot \rho(T,Y)  \cdot \rho(T,Z), \label{MP}\\
\rho_{R_{5}}(T,X,Y,V,Z)=\rho(X,Y)\rho(Y,Z)\rho(X,Z)\rho(T,X)\rho(T,Y)\rho(T,Z)\rho(X,V) \rho(Y,V) \rho(V,Z) \rho(T,V) \label{MO}
\end{eqnarray} 
\end{widetext}
In general case $|S|=m$ similarly to (\ref{MQ})-(\ref{MP}) we derive simplified formula for transitivity measure of the $m$-s member of hierarchy:
\begin{equation}
\label{transord}
\rho_{R_{m}}(X_{1},X_{2},\dots X_{m})=\prod_{i<j \le m}\rho(X_{i},X_{j})
\end{equation}
\section{Transitivity as ordering relation's property}\label{ord}
\begin{figure}
\begin{center}
\setlength{\unitlength}{0.12mm}
\begin{picture}(250,250)(0,-25)
\put(-53,50){\line(3,5){60}}
\put(67,50){\line(-3,5){60}}
\put(-53,50){\line(1,0){120}}
\put(-43,30){\small{\makebox(0,0){${ CRB}$}}}
\put(-33,170.2){\small {\makebox(0,0){${SPX}$}}}
\put(56,30){\small{\makebox(0,0){$USB$}}}
\put(7,70){\large{\makebox(0,0){${-}$}}}
\put(69,110.1){\large{\makebox(0,0){${+}$}}}
\put(-53.40,110.1){\large{\makebox(0,0){${-}$}}}
\put(7,-20){\large{\makebox(0,0){$\in V_{T}$}}}
\put(197,50){\line(3,5){60}}
\put(317,50){\line(-3,5){60}}
\put(197,50){\line(1,0){120}}
\put(207,30){\small{\makebox(0,0){${ CRB}$}}}
\put(217,170.2){\small {\makebox(0,0){${USD}$}}}
\put(306,30){\small{\makebox(0,0){$USB$}}}
\put(257,70){\large{\makebox(0,0){${-}$}}}
\put(319.1,110.1){\large{\makebox(0,0){${-}$}}}
\put(197.40,110.1){\large{\makebox(0,0){${-}$}}}
\put(257,-20){\large{\makebox(0,0){$\in V_{F}$}}}
\end{picture}
\caption{
\label{fig0}
{The three points complexes: transitive and frustrated. The $\pm$ signs correspond to the signs of the correlation coefficients.}
}
\end{center}
\end{figure}
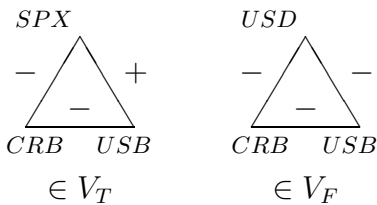

{\em Proposition}\\  
Let us consider two examples of plaquettes, one by one  from $V_{T}$ and  $V_{F}$, respectively (Fig. \ref{fig0}).
 We argue for the following hypothesis: Transitivity of ${\bf {\hat{R}_{2}}}$ is responsible for stimulating of  sectors to common direction evolution. However,  some sectors interferes with this process leading to the frustration and in this way they preserve the sectors' independence.There are two arguments for such interpretation of the frustration (or its contradiction - transitivity).\\
 The first one is direct. Let us take into account the case $V_{T}$ of Fig. \ref{fig0}. Let us estimate influence of $CRB$ on $USB$. There are two ways of entanglement: $CRB\rightarrow USB$ and $CRB\rightarrow  SPX\rightarrow USB$. Both of them push $USB$ into  opposite direction with respect to the evolution's direction of $CRB$. It is important that both ways push $USB$  into the same direction (by the direction of X's evolution we mean its increase or decrease). Therefore, the resulting effect from the both ways is at least stronger then the strongest single entanglement in this plaquette.   This result is invariant with respect to a choice of starting point in the considered plaquette. Now, let us take into account the case $V_{F}$ of  Fig. \ref{fig0}, and estimate influence of $CRB$  on $USD$. There are two ways of entanglement acting on $USD$: $CRB\rightarrow USD$ and $CRB\rightarrow  USB\rightarrow USD$. However, now they work in opposite directions. Therefore, the resulting effect is at least weaker then the strongest single entanglement in this plaquette. Also this result is invariant with respect to a choice of the starting point. Summarizing, we have shown argument that in a plaquette of the three different sectors without (with) frustration the influences of sectors between each other  become stronger (weaker). Summarizing, we distinguish an ordering entanglement in $V_{T}$, whereas in $V_{F}$ such an ordering does not exists.\\
The second argument is formal and touches the basis of the mathematics. 
Let ${\bf {\hat{R}_{2T}}},{\bf {\hat{R}_{2F}}}$ be ${\bf {\hat{R}_{2}}}$ 
constrained to $V_{T},V_{F} $, respectively. 
Due to symmetry,  reflexibility and transitivity of 
${\bf {\hat{R}_{2T}}}$ this subrelation is an equivalence relation. 
Since for the considered Intermarket the sum of the all plaquettes belonging to $V_{T}$ is equal to $S$:
\begin{equation}
\label{sum}
\bigcup_{\{X,Y,Z\}\in V_{T}}\{X,Y,Z\}=S,
\end{equation}
 the structure $(S,{\bf {\hat{R}_{2T}}})$  is  a preorder \citep{bib:foldes}.   Therefore, the transitivity is an inductor of at least a weak  kind of order in the system. \\
\section{Frustrations Hierarchy Analysis of the U.S. Intermarket's} \label{anal}
Applying (\ref{transord} ) to the Intermarket's data we have calculated the total Intermarket's transitivity measure $\rho_{\bf R_{5}}$ (see Fig.\ref{fig:rhoR5g}, Fig.\ref{fig:rhoR5d}) and the five measures
$\rho_{\bf R_{4}}$ corresponding to Subintermarkets obtained by reduction of the considered Intermarket with respect  to each its element $S$\textbackslash$\{XYZ\}$ (see Fig.\ref{fig:notXAU}- Fig.\ref{fig:notSPX}). 
\subsection{Discussion of results for 
$\rho_{\bf R_{5}}$ measure}
\begin{figure}[!]
\includegraphics[ width=8cm]{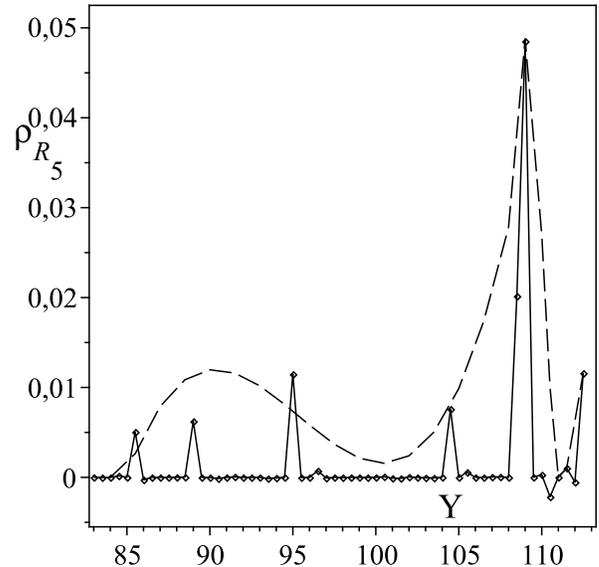}
\caption{ \label{fig:rhoR5g}The transitivity's measure ${\bf\rho_{R_{5}}}$ of $S$ v.s. $Y$ for the period 1983-2012. The vertical scale enables to read the positive values of ${\bf\rho_{R_{5}}}$, ($Y=Year-1900$).}
\end{figure}

\begin{figure}[!]
\includegraphics[ width=8cm]{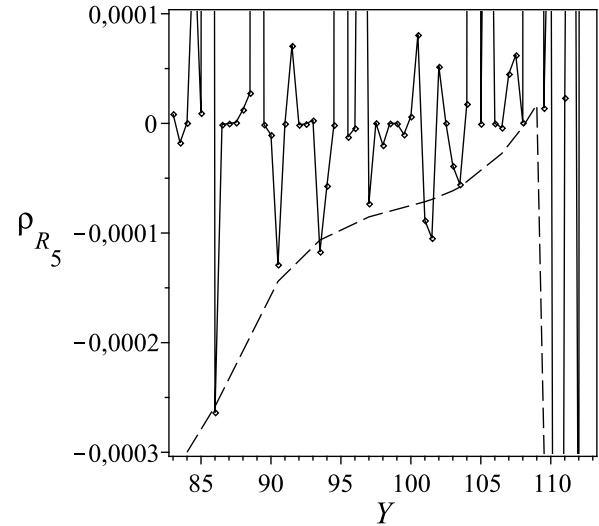}
\caption{ \label{fig:rhoR5d}The transitivity's measure ${\bf\rho_{R_{5}}}$ of $S$ v.s. $Y$ for the period 1983-2012. The vertical scale enables to read the negative values of ${\bf\rho_{R_{5}}}$ and their oscilations around $Y$ axis ($Y=Year-1900$).}
\end{figure}

The values of $\rho_{\bf R_{5}}$ are presented in the two scales. The scales of Fig.\ref{fig:rhoR5g} and Fig.\ref{fig:rhoR5d} are appropriate for analysis of positive  and  negative  $\rho_{\bf R_{5}}$, respectively. Combining both figures we see that $\rho_{\bf R_{5}}$ undergoes variations like the U.S. Business Cycles which can be described by the Brownian Motion of a Harmonic Oscillator \citep{bib:chen}-\citep{bib:Zarn}. The oscillation's amplitude for the positive direction of $\rho_{\bf R_{5}}$ is two orders greater in average then the negative one. The envelopes of positive and negative values changes according to trends presented by dashed  lines. Let us remind that $\rho_{\bf R_{5}}>0$ corresponds to transitivity, whereas $\rho_{\bf R_{5}}<0$ corresponds to frustration. From 1983 $(Y=83)$ until 2009 $(Y=109)$ the positive amplitude increases and the negative one decreases becoming positive. It means that the transitivity approaches a high value and the frustration disappears. Then from the second half of 2009 $\rho_{\bf R_{5}}$ exhibits strong fluctuations. One can see from Fig.\ref{fig:rhoR5d} that the considered system has lost  stability when the trend of frustrations got value equal to zero. On the basis of these observations we may draw the conclusion that frustration is necessary for the system's stability. Due to the strong oscillations which have appeared after the revealed boom there is a chance to get negative values for the frustration's trend and to recover  the system's stability.

The nearest future will show how the frustration analysis applied to an Intermarket is efficient for the predictions of the economic stability in long time horizon. Now (January 2013)  $\rho_{R_{5}}$ performs large amplitude oscillations into both directions ($\rho_{R_{5}}>0$ and $\rho_{R_{5}}<0$). The most probable event is that
the envelope of frustrations $\rho_{R_{5}}<0$ will approach zero by forthcoming decades and different events of U.S. Economy will generate picks of transitivity. Unlikely but a worse one would be a situation when $\rho_{R_{5}}$ will oscillate above $Y$ axis approaching  zero value asymptotically for long time.

\subsection{Discussion of results for 
$\rho_{\bf R_{4}}$ measures}

\begin{figure}[!]
\includegraphics[ width=8cm]{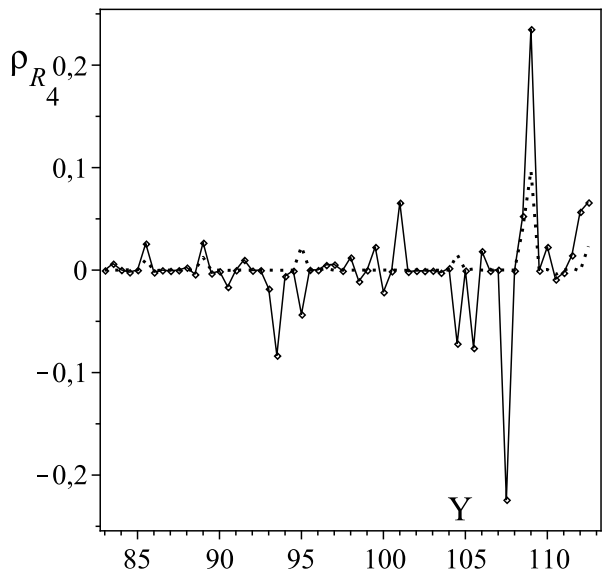}
\caption{\label{fig:notXAU} The  measure ${\bf\rho_{R_{4}}}$ of $S$\textbackslash$\{XAU\}$ v.s. $Y$ for the period 1983-2012, dots correspond to scaled $\rho_{\bf R_{5}}$, and $Y=Year-1900$.}
\end{figure}

\begin{figure}[!]
\includegraphics[ width=8cm]{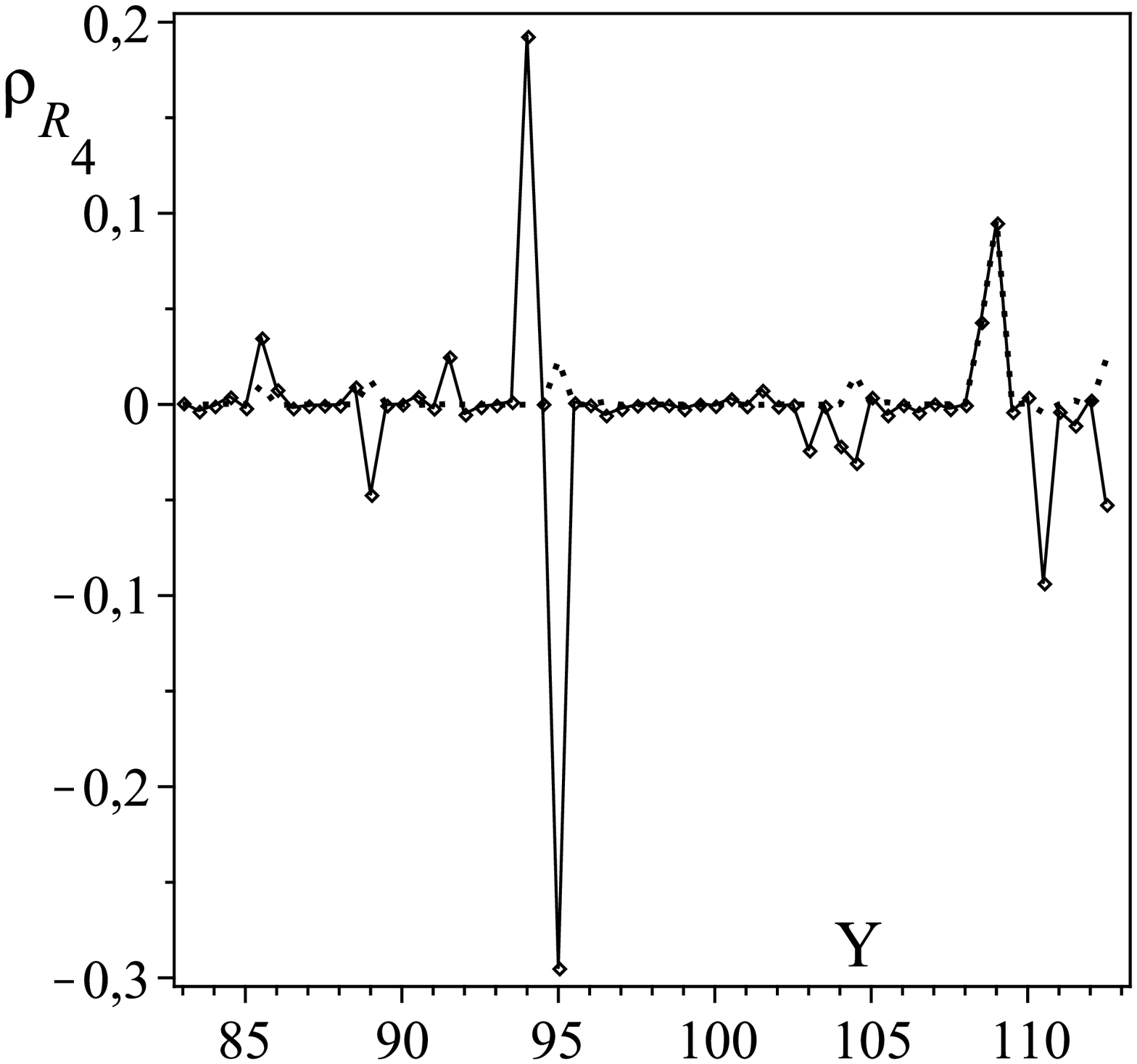}
\caption{\label{fig:notUSB}The  measure ${\bf\rho_{R_{4}}}$ of $S$\textbackslash$\{USB\}$ v.s. $Y$ for the period 1983-2012, dots correspond to scaled $\rho_{\bf R_{5}}$, and $Y=Year-1900$.}
\end{figure}

\begin{figure}[!]
\includegraphics[ width=8cm]{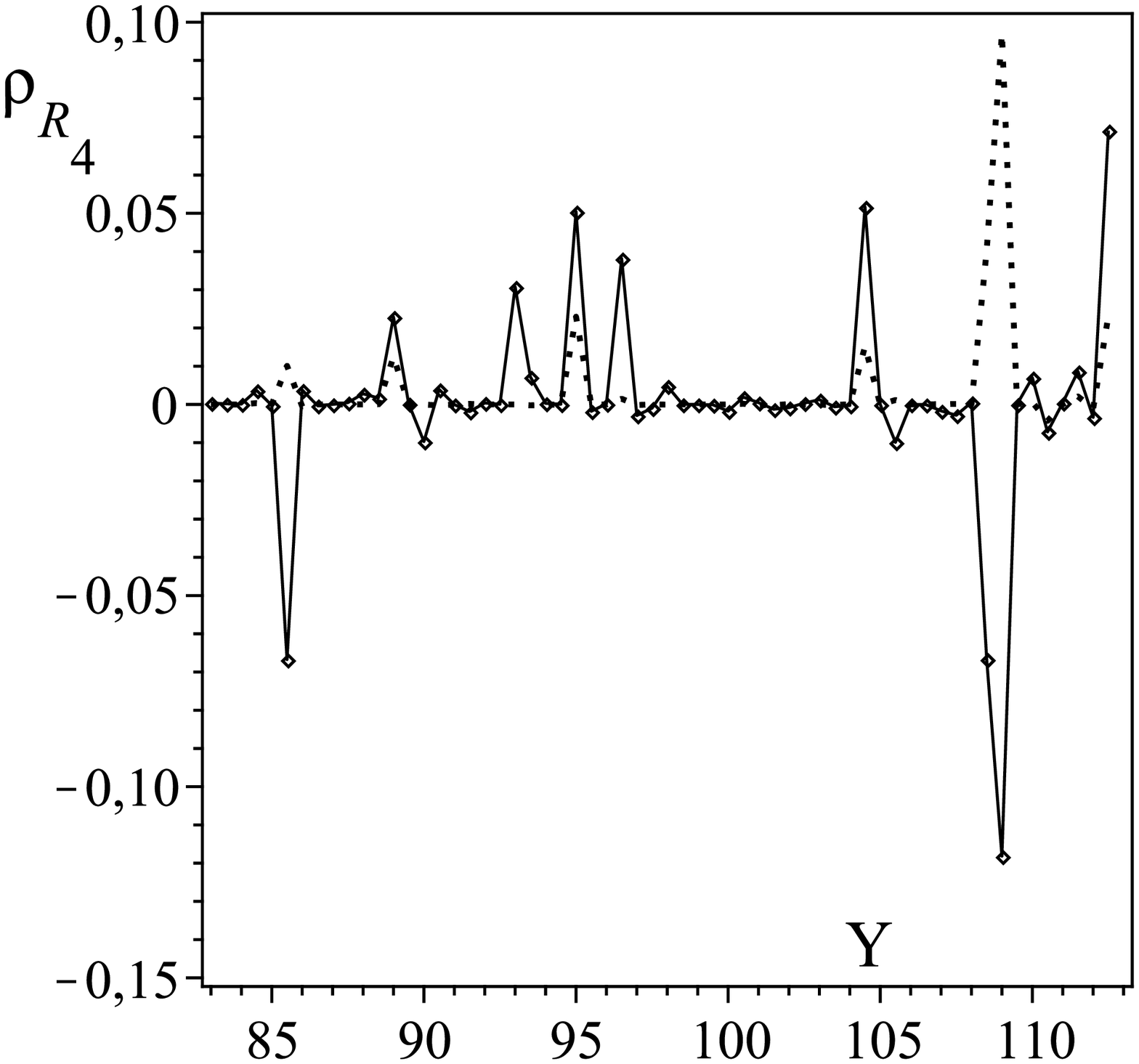}
\caption{\label{fig:notUSD} The measure ${\bf\rho_{R_{4}}}$ of $S$\textbackslash$\{USD\}$ v.s. $Y$ for the period 1983-2012, dots correspond to scaled $\rho_{\bf R_{5}}$, and $Y=Year-1900$.}
\end{figure}

\begin{figure}[!]
\includegraphics[ width=8cm]{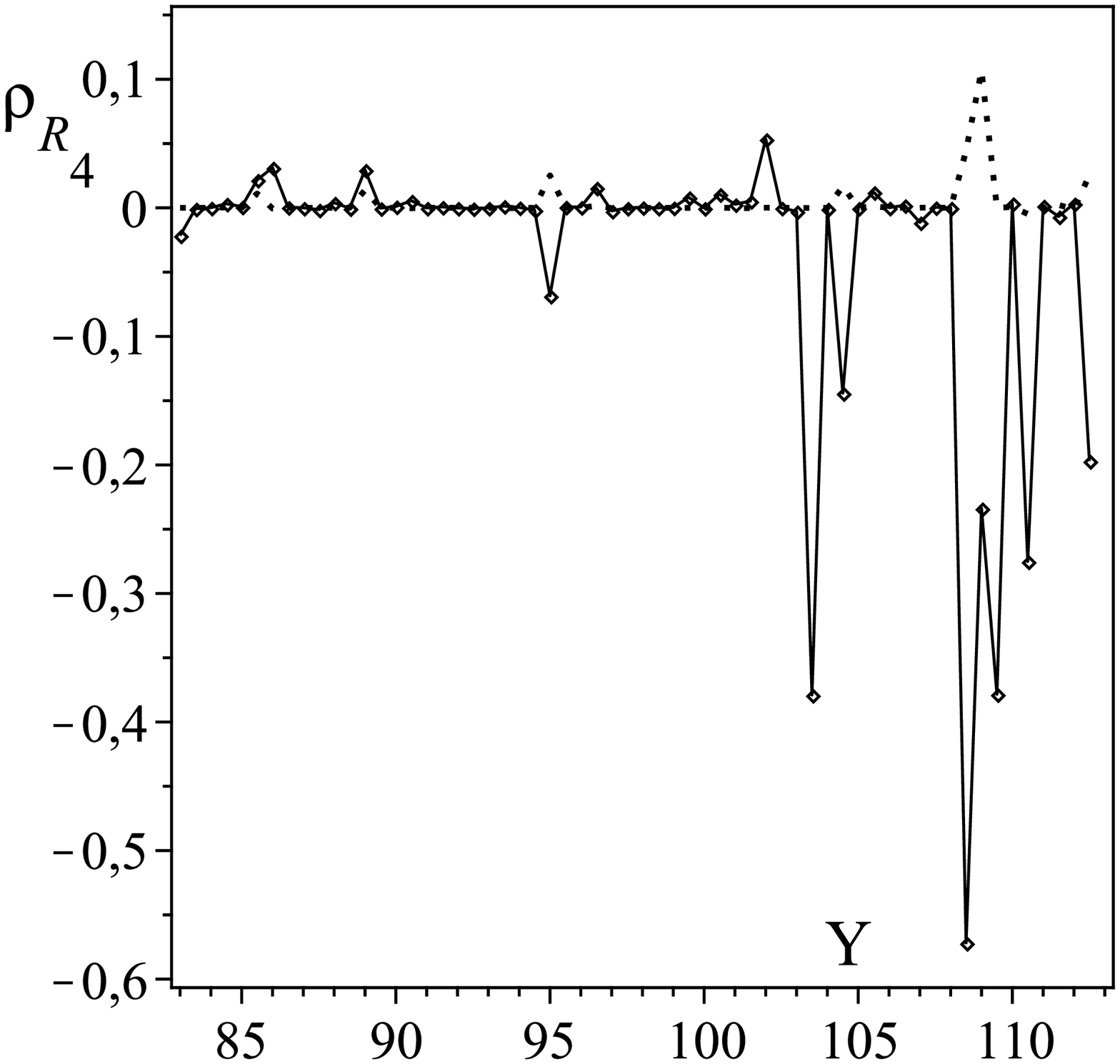}
\caption{\label{fig:notCRB} The  measure ${\bf\rho_{R_{4}}}$ of $S$\textbackslash$\{CRB\}$ v.s. $Y$ for the period 1983-2012, dots correspond to scaled $\rho_{\bf R_{5}}$, and $Y=Year-1900$.}
\end{figure}

\begin{figure}[!]
\includegraphics[ width=8cm]{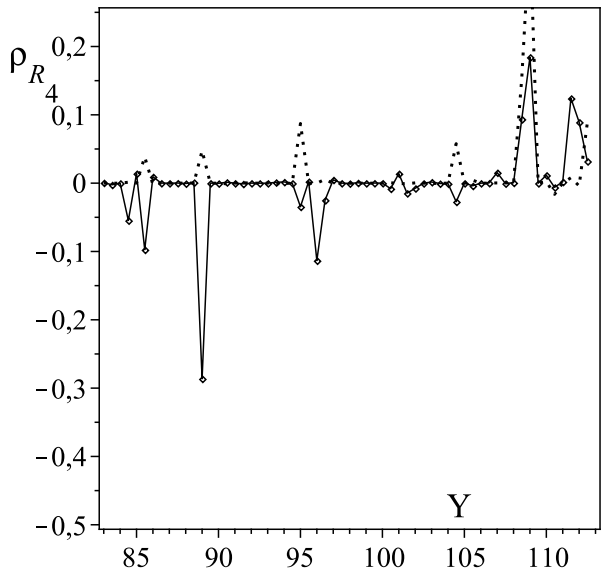}
\caption{\label{fig:notSPX} The  measure ${\bf\rho_{R_{4}}}$ of $S$\textbackslash$\{SPX\}$ v.s. $Y$ for the period 1983-2012, dots correspond to scaled $\rho_{\bf R_{5}}$,($Y=Year-1900$).}
\end{figure}

$\rho_{\bf R_{5}}$ is the highest measure of transitivity in the five elements system. This is a top of the considered hierarchy $\rho_{\bf R_{5}},\rho_{\bf R_{4}},\rho_{\bf R_{3}},\rho_{\bf R_{2}}$.
There are five measures of the $\rho_{\bf R_{4}}$ describing transitivity in a subset of four Intermarket's entities. Let us assume that by removing selected entity from the frustration analysis we receive an approximation knowledge about the influence of this Intermarket's member on the dynamics of the whole system. All the five $\rho_{\bf R_{4}}$ measures are presented in Fig.\ref{fig:notXAU}-Fig.\ref{fig:notSPX}. Comparing  $\rho_{\bf R_{4}}$ of the selected Subintermarket with $\rho_{\bf R_{5}}$ we will try to answer the question what could be the influence of the removed entity on the Intermarket's stability.\\

There are seven possible reactions of picks for removing a sector from the Intermarket. All of them are listed in TABELE \ref{1} and  their interpretations are indicated.
According to this Tabele we present the following discussion.
\begin{itemize}
\item{$S$\textbackslash$\{XAU\}$}. In the period 1983-2008  Gold has played crucial role in  Intermarket's stability. By removing  Gold from the Intermarket the high picks of transitivity have change into deep frustrations. Therefore the Gold was responsible for blocking of the frustrations. However, in the period 2009-2012 this Sector has loss this influence. 
\item{$S$\textbackslash$\{USB\}$}. In the period 1983-1994 Treasury Bonds Prices' influence was marginal. Whereas, in the period 1995-2012 $USB$ has changed transitivities into frustrations. Therefore, probably this sector among others was responsible for the frustrations' decay.  

\item{$S$\textbackslash$\{USD\}$}. The analysis of ${\bf\rho_{R_{4}}}$ shows that $USD$ is the most complicated Intermarket's sector. The pick of ${\bf\rho_{R_{4}}}$ corresponding to 1985-1986 period has been changed from frustration to transitivity. Therefore the U. S. Dollar was a creator of frustrations in this period. The pick corresponding to 1989 was invariant with respect to removing of USD from the Intermarket. Therefore for this period USD was marginal. The two next picks of ${\bf\rho_{R_{4}}}$ have appeared at 1995 and 2009. Both of them were  also invariant, however the pick from 1995 had got two sattelite picks at 1994 and 1996.5. (July of 1996). The pick at 2009 has changed from frustration into transversity whereas the last one  at  2012 reminded to be invariant. Summarizing, the invariant picks are not correlated with USD, however those which have been changed from frustration into transfersity corresponded to frustrations' creator of ${\bf\rho_{R_{5}}}$. 

\item{$S$\textbackslash$\{CRB\}$}. Influence of Commodity  on the Intermarket was a little different from the influence of another sectors.  In 1985-1986 the pick of ${\bf\rho_{R_{4}}}$ 
was invariant. The pick at 1989 has changed the  sign and became a measure of frustration. Therefore,  in this year  Commodity has  protected Intermarket against frustration. A new pick of transitivity has appeared at 1991.5.  In the period 1993.5-
1995.5 a new quality has occur. By comparing Fig. \ref{fig:notCRB} and Fig. {fig:rhoR5g} a large increase of the pick at 1995 and its little satellite at 1996.5 into gigantic  measure's values of transitivity and frustration have been observed. This can be interpreted as an active role of Commodity in creation of the frustrations in the considered period.  Next, in 2004-2005 the pair of  transitivities' picks  has been collapsed into fuzzy frustrations.Therefore, the Commodity has stopped the creation of frustrations . Finally, the events of the period 2008-2011 have shown transition of the gigantic transitivities'  into picks of transitivity and frustration. Therefore,  from 1993.5 the Commodity has played very important role in the Intermarket's stabilization.
\item{$S$\textbackslash$\{SPX\}$}. Two picks of transitivity at 1985 and 1989 have been conserved, whereas the pick of 1995 and its little satellite at 1996 have changed for frustration, moreover, the intensity of the satellite increased. Since 1997 it has started the oscillation of $\rho_{R_{4}}$ around zero characterized by increasing amplitude which was suddenly broken at 2001.5. There were not any oscillations of the measure ${\bf\rho_{R_{5}}}$ corresponding to that ones. In  the four previous cases the events after 2008 were most interesting. In the case of  $S$\textbackslash$\{SPX\}$ the gigantic picks of  transversity and frustration has been reversed in time (frustration and transversity). The five last points in Fig.\ref{fig:notSPX} suggest that the black scenario for developing of 
${\bf\rho_{R_{5}}}$ without frustrations would be possible (black scenario).   Note that in the period $1983 - 2008$ Stock and Intermarket were strongly anticorrelated,  whereas in $2009$ this relation become to be a strong correlation.  Additionaly, this event checks  the stabilizing role of frustrations.
\end{itemize}
\subsection{The finale remark}
The considered Intermarket is a subsystem immersed in the U.S. National Economy. Therefore,  the picks and trends of the transitivity as well as the frustration measures are 
related to  events of this global system. Analysis of the results presented here will be related to U.S. Events \citep{bib:USE} and published in forthcoming monograph \citep{bib:SokMor}.
 \begin{widetext}
\begin{center}
\squeezetable
\begin{table}[!t]
\renewcommand{\arraystretch}{1.3}
\caption{ Possible reactions of 
$\rho_{R_{5}}\rightarrow \rho_{R_{4}}$ for removing a sector from the Intermarket }
\label{1}
\centering
\begin{tabular}{|c||c|c|c|c|c|c|c|}
\hline
Reaction &Invariant & $F\rightarrow T$ & $T\rightarrow F$ & $F\rightarrow 0$ & $T\rightarrow 0$ & $0\rightarrow F$ & $0\rightarrow T$ \\ 
\hline
Interpretation for sector's&No active  & Frustration's&Transisivity's &Frustartion's &Transitivity's & Frustration's&Transitivity's\\
activity&  &generator &generator &generator & generator&annihilator &annihilator\\
\hline

\end{tabular}\\ \vspace{1mm}
\end{table}

\end{center}
 \end{widetext}


\end{document}